\documentclass[aps,prl,preprintnumbers,amssymb,superscriptaddress,twocolumn]{revtex4-1}
\usepackage[dvipdfmx]{graphicx}
\usepackage{bm}
\usepackage{amsmath}
\usepackage{ascmac}
\usepackage{setspace}
\usepackage{amssymb}
\usepackage{latexsym}
\usepackage{mathtools}
\usepackage{color}

\begin{document}

\title{Floquet Spinor Bose Gases} 

\author{Kazuya Fujimoto}
\affiliation{Department of Physics, University of Tokyo, 7-3-1 Hongo, Bunkyo-ku, Tokyo 113-0033, Japan}

\author{Shun Uchino}
\affiliation{Weseda Institute for Advanced Study, Waseda University, Shinjuku, Tokyo 169-8050, Japan}

\date{\today}

\begin{abstract}
We introduce a Floquet spinor Bose-Einstein condensate induced by a periodically driven quadratic Zeeman coupling whose frequency is larger than any other energy scales. By examining a spin-1 system available in ultracold atomic gases, we demonstrate that such an external driving field has great effect on the condensate through emergence of a unique spin-exchange interaction. We uncover that the ferromagnetic condensate has several unconventional stationary states and thus exhibits rich continuous phase transitions. On the other hand, the antiferromagnetic condensate is found to possess a nontrivial metastable region, which supports unusual elementary excitations and hysteresis phenomena. 

\end{abstract}

\maketitle

{\it Introduction.-} Quantum degenerate systems with multiple order parameters emerge in diverse fields of physics such as unconventional superconductors \cite{supercon1,supercon2} and superfluid $^3{\rm He}$ \cite{heli3_1,heli3_2} in condensed matter, $p$-wave superfluids in neutron stars \cite{neutron_star1,neutron_star2}, and color superconductors in quark matter \cite{color_super1,color_super2}. Due to presence of nontrivial order-parameter manifolds, such  systems are known to exhibit a variety of phase structures, low-energy excitations, and topological defects absent in single order-parameter systems including conventional $s$-wave superconductors. 

Currently, spinor Bose-Einstein condensates (BECs) realized in ultracold atomic gases offer testing grounds for examining fundamental properties of multiple order-parameter systems \cite{binary1,spinor1,spinor2,dipolar1,dipolar2}. In fact, cold-atom experiments have successfully observed rich phase structures \cite{Stenger,bookjans,jacob}, exotic topological excitations such as solitons \cite{soliton1,soliton2,soliton3,soliton4,soliton5}, skyrmions \cite{skyrmion1,skyrmion2,skyrmion3}, knots \cite{knot1,knot2}, and vortices \cite{vortex1,vortex2,vortex3}, and universal non-equilibrium dynamics \cite{universal_dyamics1,universal_dyamics2,universal_dyamics3}. 

One of the strengths in ultracold atomic gases is high controllability of experimental parameters, e.g. atom-photon interactions \cite{ultra_cold2}. Of particular interest using this controllability is Floquet engineering \cite{Floquet1, Floquet2, Floquet3, Floquet4}, where a periodically oscillating field is applied to a system and thereby generates unconventional states absent in equilibrium \cite{Floquet_state1,Floquet_state2}. In ultracold atomic gases, Floquet engineering has successfully been implemented \cite{Haldane1,Haldane2,Hofsta1,Hofsta2,Frust}, and one of the remarkable realizations is an artificial gauge field \cite{AGF1,AGF2,AGF3,AGF4}. Despite the surge of great interest in Floquet engineering, such a technology in BECs has mainly been limited to engineering of kinetic energy terms.

\begin{figure}[b]
\begin{center}
\includegraphics[keepaspectratio, width=8.5cm,clip]{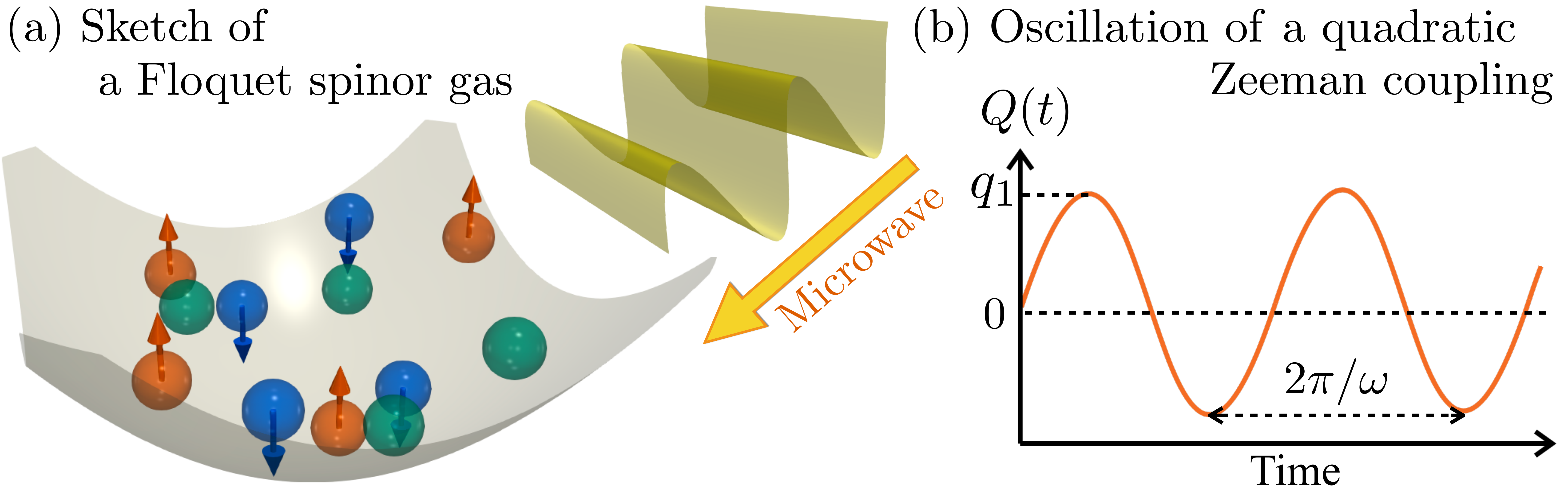}
\caption{(a) Schematic illustration of a Floquet spin-1 Bose gas induced by a microwave. Spheres with up and down arrows show particles with the magnetic sublevels $m=1$ and $-1$, and those without arrows do particles with $m=0$. (b) A modulation of microwave's amplitude introduces a time-dependent quadratic Zeeman coupling $Q(t)$ with the frequency $\omega$ and the amplitude $q_1$.
\label{fig1}} 
\end{center}
\end{figure}

In this Letter, we introduce a Floquet spinor BEC induced by a high-frequency modulation of an external field  (see Fig.~\ref{fig1}), and uncover emergence of an unconventional spin-exchange interaction. As a possible external field, we consider a microwave, which is known to cause an effective quadratic Zeeman shift in spinor systems \cite{gerbier,leslie,bookjans,jacob,guzman,tokuno}. By employing high-frequency expansion in the Floquet formalism \cite{AGF2,high_fre_exp1,high_fre_exp2}, we obtain an effective Hamiltonian having the unconventional interaction, which is in sharp contrast to the case of an artificial gauge field (note however Ref.~\cite{AGF5}). Applying the theory of a weakly-interacting Bose gas \cite{BdG}, we demonstrate that the effective Hamiltonian leads to several nontrivial stationary states and elementary excitations absent in the non-driven system. In the spinor BEC with a ferromagnetic (FM) interaction, we find emergence of the unconventional stationary states and various continuous quantum phase transitions. The most notable state is a titled broken axisymmetry (TBA) phase, which spontaneously breaks $Z_2$ symmetry despite absence of the linear Zeeman effect. In the system with an antiferromagnetic (AFM) interaction, on the other hand, we find a metastable region where two independent nonmagnetic states are stabilized simultaneously. This leads to hysteresis phenomena, which do not emerge in conventional spinor systems without the linear Zeeman effect \cite{comment}. 

{\it Theoretical model.-}
We consider spin-1 atoms in a uniform system. Under the influence of a microwave, the Hamiltonian, $\hat{H} = \hat{H}_{\rm free} + \hat{H}_{\rm int} + \hat{H}_{\rm drive}$, is given by  \cite{spinor1,spinor2}
\begin{eqnarray}
&& \hat{H}_{\rm free} = \int d\bm{r}   \sum_{m=-1}^1 \hat{\psi}_m^{\dagger}(\bm{r})  \Big (  -  \frac{\hbar^2}{2M} \nabla^2   + qm^2   \Big) \hat{\psi}_m(\bm{r}) , 
\label{free_hamil}\\
&&\hat{H}_{\rm int} = \int d\bm{r}   \Big   (   \frac{c_0}{2} :\hat{n}(\bm{r}) \hat{n}(\bm{r}): + \frac{c_2}{2} :\hat{\bm{F}}(\bm{r}) \cdot \hat{\bm{F}}(\bm{r}):  \Big), 
\label{int_hamil}
\end{eqnarray}
\begin{eqnarray}
\hat{H}_{\rm drive} = \int d\bm{r}  \sum_{m=-1}^1  Q(t) m^2  \hat{\psi}_m^{\dagger}(\bm{r}) \hat{\psi}_m(\bm{r}) := Q(t)\hat{D}, 
\label{drive_hamil}
\end{eqnarray}
where $ \hat{\psi}_m (\bm{r}) $ is the field operator of an atom with mass $M$ in a magnetic  sublevel $m$ at position $\bm{r}$, $c_0$ ($c_2$) is the spin-independent (spin-dependent) coupling, $q$ is the static quadratic Zeeman coupling, $Q(t)$ is the driving Zeeman coupling, and $::$ denotes normal ordering. The density and spin density operators are respectively defined as $\hat{n}(\bm{r}) = \sum_{m=-1}^{1} \hat{\psi}_m^{\dagger}(\bm{r}) \hat{\psi}_m(\bm{r})$ and $\hat{F}_{\mu}(\bm{r}) = \sum_{m,n=-1}^{1} \hat{\psi}_m^{\dagger}(\bm{r}) (S_{\mu})_{mn} \hat{\psi}_n(\bm{r})$ with the spin-1 matrix $(S_{\mu})_{mn}~(\mu=x,y,z)$. 

\begin{figure}[t]
\begin{center}
\includegraphics[keepaspectratio, width=8.5cm,clip]{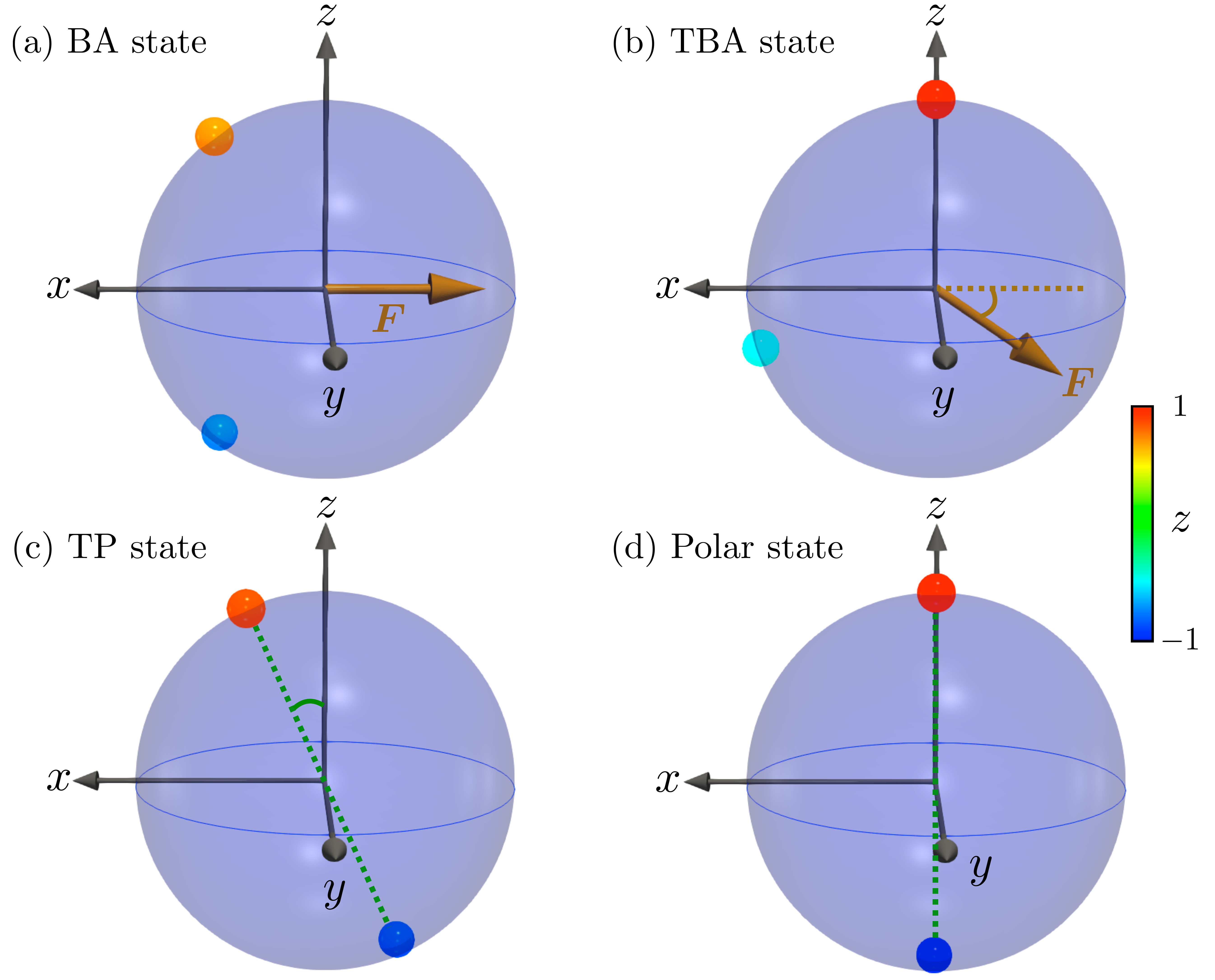}
\caption{Magnetization $\bm{F}$ and Majorana representation for (a) BA, (b) TBA, (c) TP and (d) polar states. The magnetization in the BA state is restricted in the $x$-$y$ plane, while that in the TBA state has the nonzero $z$-component. In contrast, the TP and the polar states have no magnetization. Symmetry of each phase can geometrically be visualized with the Majorana representation expressed by two spheres \cite{spinor1,spinor2,Maj}. Colors show values of $z$ for each sphere. If spheres are invariant under a transformation, a state has the corresponding symmetry. 
\label{fig2}} 
\end{center}
\end{figure}

{\it Application of high-frequency expansion.-}
We focus on the case of the periodic driving $Q(t) = q_1 {\rm cos}( \omega t)$ with the amplitude $q_1$ and the frequency $\omega$ larger than any other time scales. Such an implementation is experimentally  achievable  via tuning a microwave~\cite{shin}. By using high-frequency expansion \cite{AGF2,high_fre_exp1,high_fre_exp2}, the system is well described by the following effective (static) Hamiltonian,
\begin{eqnarray}
\hat{H}_{\rm eff} = \hat{H}_{0}  + \frac{q_1^2}{4 \hbar^2 \omega^2} \biggl[ \Big[ \hat{D}, \hat{H}_{\rm 0}  \Big],  \hat{D} \biggl] + \mathcal{O}(\omega^3),
\label{eff_hamil1} 
\end{eqnarray}
with $\hat{H}_{0} = \hat{H}_{\rm free}  + \hat{H}_{\rm int}$. Substituting Eqs. \eqref{free_hamil}$-$\eqref{drive_hamil} into Eq. \eqref{eff_hamil1}, we obtain
\begin{eqnarray}
\hat{H}_{\rm eff} = \hat{H}_{0}  - c_{\rm f} \int d\bm{r} \big( \hat{\psi}_1^{\dagger} (\bm{r}) \hat{\psi}_{-1}^{\dagger}(\bm{r}) \hat{\psi}_0(\bm{r}) \hat{\psi}_0(\bm{r}) + {\rm h.c.} \big),
\label{eff_hamil2}
\end{eqnarray}
with $c_{\rm f} = q_1^2 c_2 / \hbar^2\omega^2$. Thus, the fast driving of the quadratic Zeeman term induces the novel spin-exchange interaction originating from the commutation relation between the quadratic Zeeman term $\hat{H}_{\rm drive}$ and the spin-dependent term of $\hat{H}_{\rm int}$. We note that symmetry of the effective Hamiltonian is same as that of $\hat{H}$, which is invariant under U(1) phase rotation, U(1) spin rotation along the $z$ direction, and $Z_2$ spin rotation along the transverse direction.

{\it Stationary solutions.-}
We examine stationary states of Eq. \eqref{eff_hamil2} within the mean-field approximation where $\hat{\psi}_{m}$ are replaced by c-numbers $\Psi_{m}$.  The variational calculus in $\Psi_{m}$ leads to  the following coupled equations:
\begin{eqnarray}
\mu \Psi_1 = (q+c_2 F_z) \Psi_1 + \frac{c_2}{\sqrt{2}} F_{-} \Psi_0 - c_{\rm f} \Psi_0^2 \Psi_{-1}^*, 
\label{GP1}\\
\mu \Psi_0=  \frac{c_2}{\sqrt{2}} ( F_{+} \Psi_1 + F_{-} \Psi_{-1})  - 2c_{\rm f} \Psi_{1} \Psi_0^* \Psi_{-1}, 
\label{GP2}\\
\mu \Psi_{-1} = (q-c_2 F_z) \Psi_{-1} + \frac{c_2}{\sqrt{2}} F_{+} \Psi_0 - c_{\rm f} \Psi_0^2 \Psi_{1}^*,
\label{GP3}
\end{eqnarray}
with the magnetization $F_{\mu} = \sum_{m,n=-1}^{1} {\Psi}_m^{*} (S_{\mu})_{mn} {\Psi}_n$, $F_{\pm}=F_x \pm i F_{y}$, and the chemical potential $\mu$ including the spin-independent interaction $c_{0}n$ with a uniform density $n$. Denoting the wavefunction as $\bm{\Psi} = \sqrt{n} \bm{\eta}$  and analytically solving Eqs.~\eqref{GP1}$-$\eqref{GP3}, we find the following six stationary states. As known (inert) states, we obtain an FM state $\bm{\eta}_{\rm FM}= (1,0,0)$, a polar state $\bm{\eta}_{\rm P}= (0,1,0)$, and an AFM state $\bm{\eta}_{\rm AFM}= (1,0,1)$ \cite{spinor1,spinor2}. 
The rest  are represented as
\begin{eqnarray}
&\bm{\eta}_{\rm BA}&= (\alpha,\sqrt{1- 2\alpha^2},\alpha), \\
&\bm{\eta}_{\rm TBA}&= (\alpha - \beta/2,\sqrt{ 1- 2 \alpha^2 - \beta^2/2 } ,\alpha + \beta/2), \\
&\bm{\eta}_{\rm TP}&= (-\beta/2, \sqrt{ 1- \beta^2/2 },\beta/2 )
\end{eqnarray}
with $\alpha = \sqrt{ (c_{\rm f}n-2c_{2}n-q)/(4c_{\rm f}n-8c_{2}n) }$ and $\beta = \sqrt{ 1 + q/c_{\rm f}n }$. The state $\bm{\eta}_{\rm BA}$ represents a broken-axisymmetry (BA) state \cite{murata}, where the nonzero magnetization lies in the $x$-$y$ plane as shown in Fig.~\ref{fig2}(a), and U(1) phase and spin symmetries are spontaneously broken and the remaining symmetry is $Z_2$. Note that 
the energy is preserved regardless of rotations around the $z$ axis. 

The states $\bm{\eta}_{\rm TBA}$ and $\bm{\eta}_{\rm TP}$ are unique to the Floquet spinor BEC, and the spin configurations and the Majorana representations are shown in Fig. \ref{fig2}(b) and (c). The TBA state $\bm{\eta}_{\rm TBA}$ has the negative value of $S_{z}$ and can point to any directions in the $x$-$y$ plane in the same way as the BA state. Due to symmetry of Eq.~\eqref{eff_hamil2} about interchange between $\Psi_1$ and $\Psi_{-1}$,  $S_{z}$ can be positive, which means that $Z_2$ symmetry of the effective Hamiltonian is spontaneously broken in the TBA state. Therefore, the TBA state should be distinguished from the similar state in the ferromagnetic spinor BEC with the linear Zeeman term where $Z_2$ symmetry along the $z$ axis is absent at the Hamiltonian level. On the other hand, $\bm{\eta}_{\rm TP}$ has zero magnetization, and the Majorana representation of $\bm{\eta}_{\rm TP}$ is seen to be tilted in compared with $\bm{\eta}_{\rm P}$ as shown in Figs.~\ref{fig2}(c) and (d). Thus, we name it a tilted poler (TP) state. 

\begin{figure}[t]
\begin{center}
\includegraphics[keepaspectratio, width=8.5cm,clip]{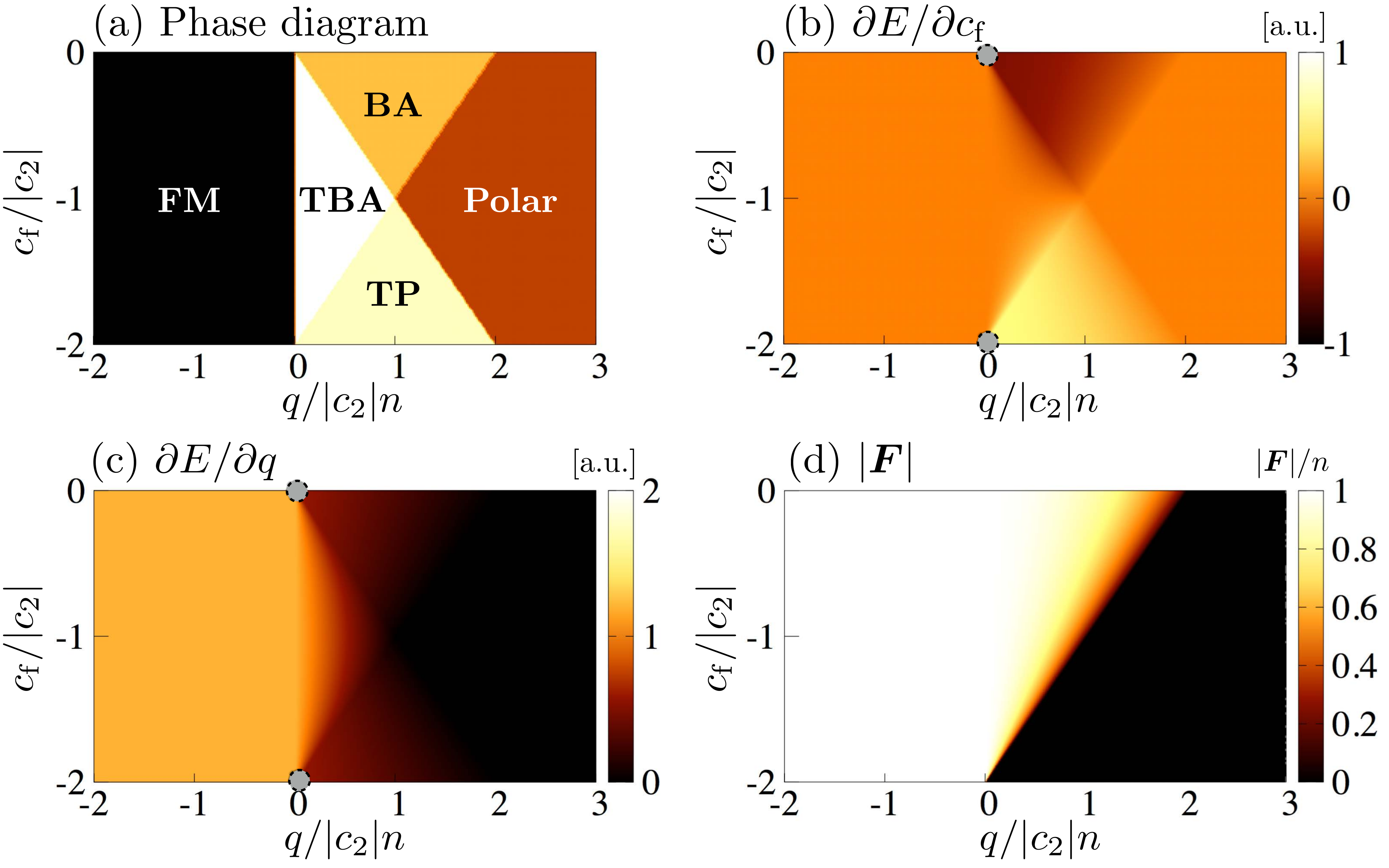}
\caption{(a) Phase diagram of the FM Floquet spin-1 BEC. While the FM, BA, and polar phases emerge in the non-driving case, the TBA and TP phases do only for nonzero $c_{\rm f}$. (b,c) Derivatives of the total energy $E$ with respect to $c_{\rm f}$ and $q$, respectively. The gray circles indicate first-order phase transition points. All other phase transitions are second order. (d) Parameter dependence of the spin amplitude $|\bm{F}|$ numerically obtained by imaginary-time evolutions of Eqs.~\eqref{GP1}--\eqref{GP3} with random initial noises. Absence of a noisy region in $|\bm{F}|$ implies that the system does not possess metastable states up to the mean-field theory. 
\label{fig3}} 
\end{center}
\end{figure}

We can obtain phase diagrams by evaluating energies of these stationary states. Figure \ref{fig3}(a) is the result for the FM interaction case ($c_2<0$), which contains the nontrivial states, $\bm{\eta}_{\rm TBA}$ and $\bm{\eta}_{\rm TP}$. We can also discuss orders of the transitions between these different phases by calculating two derivatives of total energy $\partial E(q,c_{\rm f}) / \partial c_{\rm f}$ and $\partial E(q,c_{\rm f}) / \partial q$. As shown in Figs. \ref{fig3}(b) and (c), we find that for $0<c_{\rm f}/|c_2|<2$, all transitions become second order. This is in sharp contrast to the non-driving case ($c_{\rm f}=0$) where a second-order transition is achieved only at the boundary between the BA and the polar phases.

\begin{figure}[t]
\begin{center}
\includegraphics[keepaspectratio, width=8.5cm,clip]{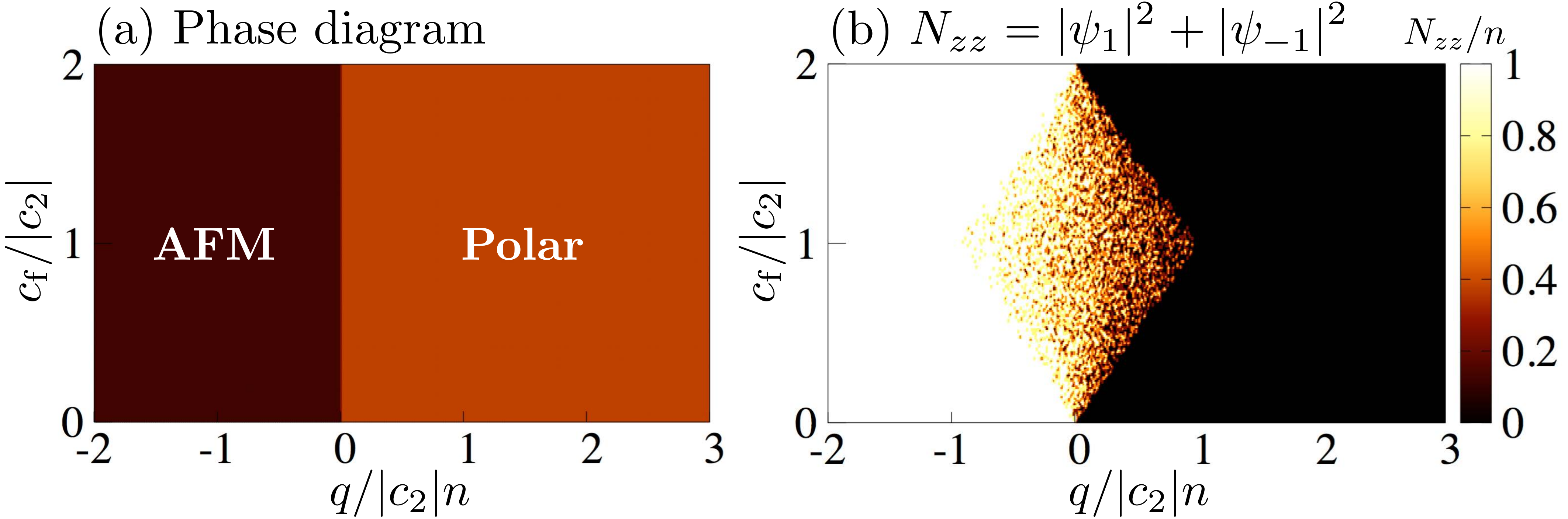}
\caption{ (a) Phase diagram of the AFM Floquet spin-1 BEC. When $q$ is positive (negative), the polar (AFM) state has  the lowest energy. (b) Distribution of nematic tensor $N_{zz}=|\Psi_{1}|^2 + |\Psi_{-1}|^2$ numerically obtained by the imaginary-time evolutions of Eqs.~\eqref{GP1}--\eqref{GP3} with random initial noises. In the noisy region, $N_{zz}$ has unity or zero. 
This implies that the AFM (polar) state for positive (negative) $q$ can be metastable in the noisy region.
\label{fig4}} 
\end{center}
\end{figure}

Figure \ref{fig4}(a) is the result for the AFM interaction case ($c_2>0$), where the transition line between the AFM and polar phases is independent of $c_{\rm f}$ and is first order. 
As discussed below, nontrivial metastable states are induced due to the periodically driving field. 

{\it Metastable states for $c_2>0$.-}
To unveil existence of the metastable states, we first perform numerical simulations of the imaginary time-step method with random initial states. The results in the FM and AFM interaction cases are respectively shown  in Figs.~\ref{fig3}(d) and \ref{fig4}(b). In the AFM interaction case, the nematic tensor ($N_{zz}=|\Psi_1|^2+|\Psi_{-1}|^2$) shows the noisy behavior, which indicates presence of metastable states. In contrast, such a noisy behavior does not show up in the FM interaction case.

We analytically understand the emergence of the metastable states for $c_2>0$ by looking at elementary excitations for $\bm{\eta}_{\rm P}$ and $\bm{\eta}_{\rm AFM}$ in the region $|q/c_2 n| \leq 1$. By employing the Bogoliubov theory \cite{spinor1,spinor2} incorporating fluctuations of $\Psi_{m}$ up to second order, we obtain the Bogoliubov Hamiltonian in each phase from Eq.~\eqref{eff_hamil2}. We diagonalize it and obtain three independent excitations known as the Bogoliubov modes. One of them is a phonon mode related to U(1) phase symmetry and common to both the AFM and the polar phases, where the excitation energy is positive definite. On the other hand, the other modes depend on symmetry of each phase. In the AFM phase, the remaining modes are given by
\begin{eqnarray}
&& E_{\rm AFM,S_{xy}}(k)  = \sqrt{ ( \epsilon_{k}  - q + c_2 n )^2 - ( c_2 n - c_{\rm f} n  )^2 },\\
&& E_{\rm AFM,S_{z}}(k)   = \sqrt{ \epsilon_{k} ( \epsilon_{k} + 2c_2 n) },
\end{eqnarray}
where $\epsilon_{k} = \hbar^2 k^2/2M$ is single particle kinetic energy, and $E_{\rm AFM,S_{xy}}(k)$ and $E_{\rm AFM,S_{z}}(k)$ denote excitations on transverse spin
and longitudinal spin fluctuations, respectively. As usual, all the Bogoliubov modes take positive values for  $q \leq 0$, where the AFM state is the lowest-energy state. What is surprising here is that these modes are positive even for the positive $q$ regime if the following inequalities are satisfied:  
\begin{eqnarray}
c_{\rm f} n < -q + 2 |c_2| n,~~~c_{\rm f}n > q ~~~ (q>0).
\label{meta1}
\end{eqnarray}
Thus, the AFM state can still be stable for $q>0$, though it is not lowest-energy state. On the other hand, in the polar phase, we obtain two degenerate modes for the transverse spin fluctuations whose excitation energy is given by
\begin{eqnarray}
E_{\rm P,S}(k) = \sqrt{ ( \epsilon_{k}  + q + c_2 n )^2 - ( c_2 n - c_{\rm f} n  )^2 }.
\end{eqnarray}
As before, for $q \geq 0$, the polar state being the lowest-energy state is stable. In addition, it can be stable even for a negative $q$ regime satisfying  
\begin{eqnarray}
c_{\rm f} n < q + 2 |c_2| n,~~~c_{\rm f}n > -q ~~~ (q<0).
\label{meta2}
\end{eqnarray}
The unusual stable region determined by Eqs.~\eqref{meta1} and \eqref{meta2} means that the AFM and polar states can exist as metastable states, and the noisy region in Fig.~\ref{fig4} (b) is indeed identical to it \cite{comment_meta}. Thus, we expect hysteresis loops when we adiabatically change $q_1$ and $\omega$ along a closed loop crossing the first-order transition.

We can discuss emergence of the metastable states from the perspective of symmetry of Eq.~\eqref{eff_hamil2}. In the absence of $c_{\rm f}$, the transition between the AFM and the polar phases is known to be first order without a metastable state \cite{spinor1,spinor2}, which is considered to be attributed to the restoration of SO(3) spin rotation symmetry at $q=0$ \cite{meta5}. However, the nontrivial spin-exchange interaction proportional to $c_{\rm f}$ breaks SO(3) rotational symmetry even at  $q=0$, so that it can stabilize the metastable states.

{\it Critical behaviors for $c_2<0$.-} 
We investiagte critical behaviors  near the second-order phase transitions in Fig.~\ref{fig3}(a). In particular, we focus on (A) BA-polar, (B) BA-TBA, and (C) FM-TBA phase transitions, since these transitions emerge in $|c_{\rm f}/c_{2}|<1$, where the effective Hamiltonian approach works well. For convenience, we introduce dimensionless variables $x=q/|c_{2}|n$ and $y=c_{\rm f}/|c_{2}|$ and denote the parameter space as $\bm{R}=(x,y)$.

\begin{figure}[t]
\begin{center}
\includegraphics[keepaspectratio, width=8.5cm,clip]{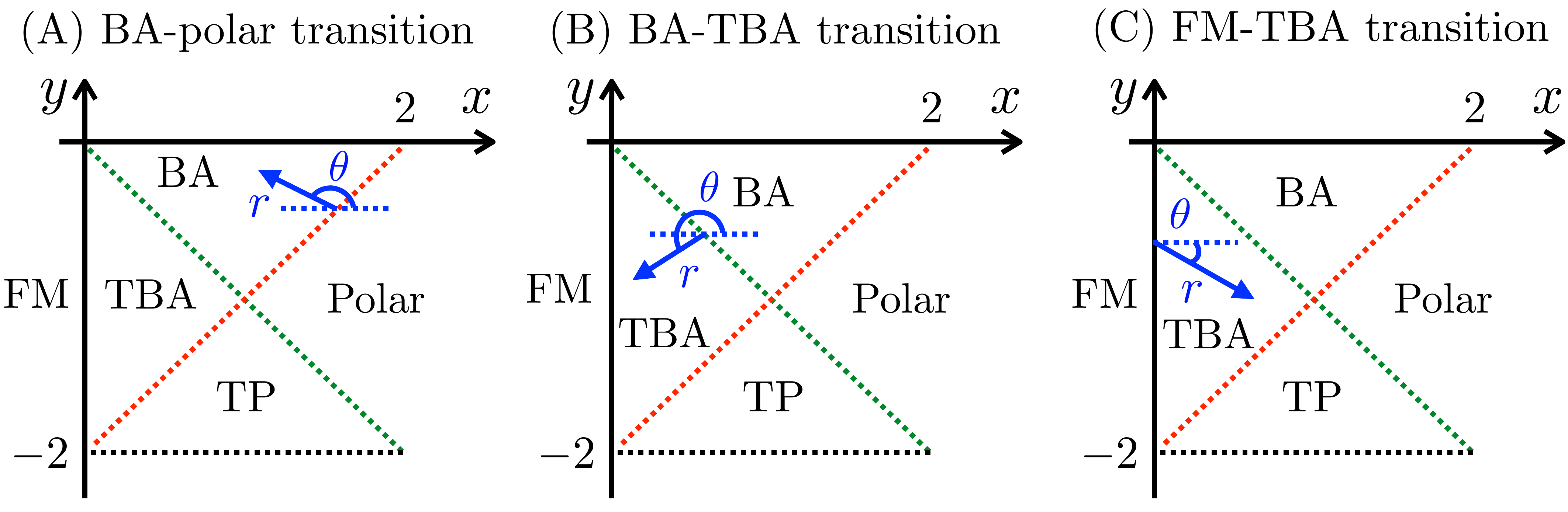}
\caption{Polar coordinates ($r$,$\theta$) near the phase boundaries: (A) BA-polar, (B) BA-TBA, and (C)  FM-TBA transitions. 
\label{fig5}} 
\end{center}
\end{figure}

First, we consider the case (A), where the phase boundary is given by $\bm {R}_{0}=(x_0,x_0-2)~(1<x_0 \leq2)$. As shown in Fig.~\ref{fig5}(A),
a point around the phase boundary is specified as $\bm {R} = ( x_0 + r{\rm cos}\theta, x_0 - 2 + r{\rm sin}\theta )$ with $r \ll 1$,
where $r$ and $\theta$ are the polar coordinates.
Then, longitudinal and transverse magnetizations in the BA phase are expressed as
\begin{eqnarray}
 F_z = 0, ~~~  F_{\perp} \propto r^{1/2},    
\end{eqnarray}
while magnetizations in the polar phase vanish. As first pointed out in Ref. \cite{XY}, such critical behaviors are same as those of the $XY$ model. 

In the case (B),  the transition point is denoted as $\bm {R}_{0}=(x_0,-x_0)~(0<x_0<1)$ as shown in Fig.~\ref{fig5}(B).
A point around the phase boundary is specified by $\bm {R}=( x_0 + r{\rm cos}\theta, -x_0 + r{\rm sin}\theta)$ with $r \ll 1$. 
Then, longitudinal and transverse magnetizations in the TBA phase are expressed as
\begin{eqnarray}
 F_{z} \propto r^{1/2}, ~~~  F_{\perp} - F_{0} \propto r, 
\end{eqnarray}
with the transverse magnetization at $\bm {R}_{0}$, $F_{0}$.
In terms of symmetry, what is essential here is that $Z_2$ spin rotational symmetry is spontaneously broken in the TBA phase while it is unbroken in the BA phase. Indeed, the leading variance on the magnetizations is  the longitudinal one directly related to $Z_2$ spin rotational symmetry, and the transition type corresponds to the Ising model.

Finally,  in the case (C), the transition point is denoted as $\bm {R}_{0}=(0,y_0)~(-2<y_0<0)$. A point around the phase boundary is specified by $\bm {R}=(r{\rm cos}\theta, y_0 + r{\rm sin} \theta)$ as shown in Fig.~\ref{fig5}(C). The magnetization behaviors in the TBA phase are expressed  as
\begin{eqnarray}
 F_{z}-n \propto r, ~~~  F_{\perp} \propto r^{1/2},
\end{eqnarray}
while in the FM phase the longitudinal magnetization coincides with $n$ and the transverse magnetization vanishes. In this transition, U(1) spin rotational symmetry is of importance since it is broken (unbroken) in the TBA (FM) phase. Therefore, the 
$XY$-type transition is expected. In addition, by using the Bogoliubov theory, the dynamical critical exponent $z$ is obtained as $z=2$. Thus, the universality corresponds to that of a single-component dilute Bose gas \cite{sachdev}.

{\it Discussion.-}
We now discuss the validity of the high-frequency expansion. According to previous literature \cite{F1,F2,F3,F4}, when a system is locally bounded, $\hbar \omega$ must be larger than the local energy. While our system is not locally bounded, we may introduce the local chemical potential $\mu \sim c_{0} n$ as an effective cutoff. Thus, $\mu < \hbar \omega$ is a necessary condition for the validity. In addition, $|c_{\rm f}/c_{2}|= |q_1/\hbar \omega |^2<1$ is required to neglect higher-order effects. Under these conditions, the fundamental properties in Floquet spinor BECs discussed above may be observed. 

To realize an FM Floquet spin-1 BEC, we can consider $^{87}$Rb and $^{7}$Li as possible atomic species. Then, observation of the TBA state can be an experimental signature of the Floquet spinor BEC. The TBA state itself can be confirmed by emergence of a spin domain along the longitudinal direction, which is absent in the non-driving case for $q>0$ \cite{chapman,saito,spinor1,spinor2}. In addition, we can use $^{23}$Na to realize an AFM Floquet spin-1 BEC. Then, the Floquet property is identified as presence of the metastable state, which can be measured through absorption images of magnetic sublevels in different times \cite{meta1,meta4}. For example, we consider an experiment in which the polar state is initially prepared at $q<0$. In the absence of the driving, such a state is unstable against the dynamical instability and the $m=\pm1$ components turn to grow up as a function of time. Namely, presence of metastability is proved as robustness of the $m=0$ component by the driving.

{\it Conclusion.-}
We have theoretically studied a spin-1 Bose gas under the periodically oscillating quadratic Zeeman coupling using the high-frequency expansion and the Bogoliubov theory. In the FM interaction case, we have found the emergence of the unconventional TBA and the TP states in addition to the known magnetic phases and the rich second-order phase transitions. On the other hand, the systems with the AFM interaction have the metastable states stabilized by the unusual Bogoliubov excitations, which can lead to hysteresis phenomena that do not emerges in undriven spinor BECs. 

This work on Floquet engineering that generates the unconventional interaction in spinor BECs paves the way towards exploring exotic states of matter in quantum fluids with internal degrees of freedom. For instance, spin-$f$ Bose gases possess $f$-independent spin-exchange interactions and thus the Floquet engineering generates the corresponding number of new interactions leading to further nontrivial phases. In the spin-2 case, the engineering may lead to an observation of non-abelian states such as a cyclic state \cite{cyclic1,cyclic2}, which has yet to be observed. An experimental observation of exotic non-abelian topological defects \cite{non_ab1,non_ab2,non_ab3} may also be discussed. 

\begin{acknowledgments}
We would like to thank Y. Shin and M. Ueda for comments in early stage of this work, and also thank S. Furukawa, R. Hamazaki, S. Higashikawa, M. Nakagawa, and M. Sato for fruitful discussions. K. F. is supported by JSPS fellowship (JSPS KAKENHI Grant No. JP16J01683).
S. U. is supported by JSPS KAKENHI Grant No. JP17K14366.
\end{acknowledgments}

\thispagestyle{myheadings}


\begin{thebibliography}{99}
\bibitem{comment} There can exist metastable states in presence of the linear Zeeman effect breaking $Z_2$ symmetry \cite{meta1,meta2,meta3,meta4}.
\bibitem{comment_meta} In the region $|q/c_2n|>1$, we also obtain spectra for Bogoliubov excitations and find no metastable states.
\end{thebibliography}
\end{document}